\documentclass[manuscript]{acmart}

\let\equiv=\Leftrightarrow 

\makeatletter
\@ifundefined{lhs2tex.lhs2tex.sty.read}%
  {\@namedef{lhs2tex.lhs2tex.sty.read}{}%
   \newcommand\SkipToFmtEnd{}%
   \newcommand\EndFmtInput{}%
   \long\def\SkipToFmtEnd#1\EndFmtInput{}%
  }\SkipToFmtEnd

\newcommand\ReadOnlyOnce[1]{\@ifundefined{#1}{\@namedef{#1}{}}\SkipToFmtEnd}
\usepackage{amstext}
\usepackage{amssymb}
\usepackage{stmaryrd}
\DeclareFontFamily{OT1}{cmtex}{}
\DeclareFontShape{OT1}{cmtex}{m}{n}
  {<5><6><7><8>cmtex8
   <9>cmtex9
   <10><10.95><12><14.4><17.28><20.74><24.88>cmtex10}{}
\DeclareFontShape{OT1}{cmtex}{m}{it}
  {<-> ssub * cmtt/m/it}{}

\DeclareFontShape{OT1}{cmtt}{bx}{n}
  {<5><6><7><8>cmtt8
   <9>cmbtt9
   <10><10.95><12><14.4><17.28><20.74><24.88>cmbtt10}{}
\DeclareFontShape{OT1}{cmtex}{bx}{n}
  {<-> ssub * cmtt/bx/n}{}

\newcommand{\Conid}[1]{\mathit{#1}}
\newcommand{\Varid}[1]{\mathit{#1}}
\newcommand{\anonymous}{\kern0.06em \vbox{\hrule\@width.5em}}
\newcommand{\plus}{\mathbin{+\!\!\!+}}

\renewcommand{\leq}{\leqslant}
\renewcommand{\geq}{\geqslant}
\usepackage{polytable}

\@ifundefined{mathindent}%
  {\newdimen\mathindent\mathindent\leftmargini}%
  {}%

\def\resethooks{%
  \global\let\SaveRestoreHook\empty
  \global\let\ColumnHook\empty}
\newcommand*{\savecolumns}[1][default]%
  {\g@addto@macro\SaveRestoreHook{\savecolumns[#1]}}
\newcommand*{\restorecolumns}[1][default]%
  {\g@addto@macro\SaveRestoreHook{\restorecolumns[#1]}}
\newcommand*{\aligncolumn}[2]%
  {\g@addto@macro\ColumnHook{\column{#1}{#2}}}

\resethooks

\newcommand{\onelinecommentchars}{\quad-{}- }
\newcommand{\commentbeginchars}{\enskip\{-}
\newcommand{\commentendchars}{-\}\enskip}

\newcommand{\visiblecomments}{%
  \let\onelinecomment=\onelinecommentchars
  \let\commentbegin=\commentbeginchars
  \let\commentend=\commentendchars}

\newcommand{\invisiblecomments}{%
  \let\onelinecomment=\empty
  \let\commentbegin=\empty
  \let\commentend=\empty}

\visiblecomments

\newlength{\blanklineskip}
\newcommand{\hsindent}[1]{\quad}
\let\hspre\empty
\let\hspost\empty

\EndFmtInput
\makeatother
\ReadOnlyOnce{polycode.fmt}%
\makeatletter

\newcommand{\hsnewpar}[1]%
  {{\parskip=0pt\parindent=0pt\par\vskip #1\noindent}}

\newcommand{\hscodestyle}{}

\newcommand{\sethscode}[1]%
  {\expandafter\let\expandafter\hscode\csname #1\endcsname
   \expandafter\let\expandafter\endhscode\csname end#1\endcsname}

  {\par\noindent
   \advance\leftskip\mathindent
   \hscodestyle
   \let\\=\@normalcr
   \let\hspre\(\let\hspost\)%
   \pboxed}%
  {\endpboxed\)%
   \par\noindent
   \ignorespacesafterend}

  {\hsnewpar\abovedisplayskip
   \advance\leftskip\mathindent
   \hscodestyle
   \let\hspre\(\let\hspost\)%
   \pboxed}%
  {\endpboxed%
   \hsnewpar\belowdisplayskip
   \ignorespacesafterend}

  {\hsnewpar\abovedisplayskip
   \advance\leftskip\mathindent
   \hscodestyle
   \let\\=\@normalcr
   \(\pboxed}%
  {\endpboxed\)%
   \hsnewpar\belowdisplayskip
   \ignorespacesafterend}

\newcommand{\plainhs}{\sethscode{plainhscode}}

\plainhs

  {\hsnewpar\abovedisplayskip
   \advance\leftskip\mathindent
   \hscodestyle
   \let\\=\@normalcr
   \(\parray}%
  {\endparray\)%
   \hsnewpar\belowdisplayskip
   \ignorespacesafterend}

  {\parray}{\endparray}

  {\(\parray}{\endparray\)}

\def\codeframewidth{\arrayrulewidth}
\RequirePackage{calc}

  {\parskip=\abovedisplayskip\par\noindent
   \hscodestyle
   \arrayrulewidth=\codeframewidth
   \tabular{@{}|p{\linewidth-2\arraycolsep-2\arrayrulewidth-2pt}|@{}}%
   \hline\framedhslinecorrect\\{-1.5ex}%
   \let\endoflinesave=\\
   \let\\=\@normalcr
   \(\pboxed}%
  {\endpboxed\)%
   \framedhslinecorrect\endoflinesave{.5ex}\hline
   \endtabular
   \parskip=\belowdisplayskip\par\noindent
   \ignorespacesafterend}

\newcommand{\framedhslinecorrect}[2]%
  {#1[#2]}

  {\(\def\column##1##2{}%
   \let\>\undefined\let\<\undefined\let\\\undefined
   \newcommand\>[1][]{}\newcommand\<[1][]{}\newcommand\\[1][]{}%
   \def\fromto##1##2##3{##3}%
   }{\) }%

  {\let\orighscode=\hscode
   \let\origendhscode=\endhscode
   \def\endhscode{\def\hscode{\endgroup\def\@currenvir{hscode}\\}\begingroup}
   \orighscode\def\hscode{\endgroup\def\@currenvir{hscode}}}%
  {\origendhscode
   \global\let\hscode=\orighscode
   \global\let\endhscode=\origendhscode}%

\makeatother
\EndFmtInput
\def\rcb#1#2#3#4{\def\nothing{}\def\range{#3}\mathopen{\langle}#1 \ #2 \ \ifx\range\nothing::\else: \ #3 :\fi \ #4\mathclose{\rangle}}
\newenvironment{lcbr}{\left\{\begin{array}{l}}{\end{array}\right.}
\def\rcbnaked#1#2#3#4{\def\nothing{}\def\range{#3}#1 \ #2 \ \ifx\range\nothing::\else: \ #3 :\fi \ #4}
\def\eqnnewpage{\end{eqnarray*}\newpage~\vskip-2em\begin{eqnarray*}&& }
\def\comp{ \mathbin{\cdot} }
\def\start{&&}
\def\more{\\&&}

\def\just#1#2{\\ &#1& \rule{2em}{0pt} \{ \mbox{\rule[-.7em]{0pt}{1.8em} \small #2 \/} \} \nonumber\\ && }

\AtBeginDocument{%
  }

\setcopyright{acmlicensed}
\copyrightyear{2024}
\acmMonth{6}
\acmYear{2024}
\acmDOI{--.--}

\acmISBN{978-1-4503-XXXX-X/18/06}

\begin{document}

\title{Haskell meets Evariste}

\author{Paulo R.\ Pereira}
\affiliation{%
  \institution{Checkmarx}
  \city{Braga}
  \country{Portugal}}
\email{prpereira.hs@gmail.com}

\author{Jos\'e N.\ Oliveira}
\affiliation{%
  \institution{HASLab/ U.Minho \& INESC TEC}
  \city{Braga}
  \country{Portugal}}
\email{jno@di.uminho.pt}


\begin{abstract}
Since its birth as a new scientific body of knowledge in the late 1950s,
computer programming has become a fundamental skill needed in many other
disciplines. However, programming is not easy, it is prone to errors and
code re-use is key for productivity. This calls for high-quality documentation
in software libraries, which is quite often not the case. Taking a few Haskell
functions available from the Hackage repository as case-studies, and comparing
their descriptions with similar functions in other languages, this
paper shows how clarity and good conceptual design can be achieved by following
a so-called \emph{easy-hard-split} formal strategy that is quite general
and productive, even if used informally. This strategy is easy to
use in functional programming and can be applied to both program
analysis and synthesis.
\end{abstract}




\maketitle

\section{Introduction}

The essence of programming lies in translating human intent into machine-executable
instructions, a task demanding clarity and unambiguous specification.
Formal specifications \cite{CW96}, which are mathematical in nature, are best at removing
ambiguity but, for large problems, they become too complex and inaccessible to the
average programmer.

Compositionality and reuse of (smaller) components made available from software
libraries come to the rescue, as functional programmers very well know. However,
how can programmers be sure about what the functions they import \emph{exactly do}?
If formal specification is not used, they will have to rely on textual descriptions,
the quality of which may vary significantly. As in any other branch of engineering,
achieving technical accuracy and high
quality readership in software documentation is very important. But this is becoming a problem, as the proficiency
of today's students (tomorrow's professional programmers) in maths and in writing and reading
is not at the required level.
Mathphobia can be overcome by novel strategies to teaching programming
such as Carroll Morgan's \emph{go formal informally} approach \cite{Mo24}.
But it will fail if students end up not understanding written texts either.

The dual of this problem happens in "forward engineering" programming, e.g.\
in so-called \emph{requirement analysis} \cite{WB13}. Here again, having
clear ideas is what matters, and relying on simple but effective \emph{concepts}
seems to be essential to good design \cite{Ja21}.

In this paper we address both sides of this problem by bridging the gap
between formal and informal description for a class of programs
whose specifications fit into a particular, but relatively broad \emph{specification pattern}.
Our analysis will focus on functional programming and, in particular, on some list manipulating 
functions available from the Hackage repository and similar repositories for other languages.

This paper is organized as follows: after a brief motivation (section \ref{sec:motivation}),
sections \ref{sec:240531f} and \ref{sec:takeWhile} address the main idea of the paper using 
the \ensuremath{\Varid{takeWhile}} function as example of a so-called \emph{easy-hard-split} functional design pattern.
The following sections generalize this example and show how the approach is formally supported.
Finally, sections \ref{sec:240603b} and \ref{sec:summary} give a brief discussion on 
why clarity and a solid basis contribute to better software understanding.

\section{Motivation} \label{sec:motivation} 
Let us analyze how some standard and intuitive functions are documented in
some of the most well-known programming languages.
We start by taking as example the documentation concerning a well-known function
--- \ensuremath{\Varid{zip}} --- as it is given in different software libraries. Its textual
specification in {Swift}, for instance, is:
\begin{quote}\em
	Creates a sequence of pairs built out of two underlying sequences.
        \cite{apple-swift}
\end{quote}
This description gives a lot of freedom of interpretation. For instance,
\ensuremath{\Varid{zip}\;[\mskip1.5mu \mathrm{1},\mathrm{2},\mathrm{3}\mskip1.5mu]\;[\mskip1.5mu \Varid{a},\Varid{b}\mskip1.5mu]\mathrel{=}[\mskip1.5mu (\mathrm{1},\Varid{b}),(\mathrm{3},\Varid{a}),(\mathrm{3},\Varid{b})\mskip1.5mu]} gives a sequence of pairs built
out of sequences \ensuremath{[\mskip1.5mu \mathrm{1},\mathrm{2},\mathrm{3}\mskip1.5mu]} and \ensuremath{[\mskip1.5mu \Varid{a},\Varid{b}\mskip1.5mu]}, \ensuremath{[\mskip1.5mu (\mathrm{2},\Varid{a}),(\mathrm{1},\Varid{a})\mskip1.5mu]} being equally acceptable.
Formally, one would say that \ensuremath{\Varid{zip}} in Swift is a \emph{relation}, not a function,
as it offers multiple acceptable outputs for the same input.
The description of \ensuremath{\Varid{zip}} given in the F\# documentation is essentially the same as Swift's:
\begin{quote}\em
	{List.zip} takes two lists of single elements and produces a single list of tuple pairs.
        \cite{ms-f-sharp}
\end{quote}

The ``same'' function in {Elixir}'s Enum library is described as follows:
\begin{quote}\em
	Zips corresponding elements from a finite collection of enumerables into a list of tuples \cite{elixir/1.12}.
\end{quote}
Under a reasonable interpretation of the adjective ``corresponding", Elixir's
\ensuremath{\Varid{zip}} gives less freedom of interpretation than Swift's and F\#'s, ruling out all tuples
in \ensuremath{[\mskip1.5mu (\mathrm{1},\Varid{b}),(\mathrm{3},\Varid{a}),(\mathrm{3},\Varid{b})\mskip1.5mu]}, for instance. But \ensuremath{[\mskip1.5mu (\mathrm{1},\Varid{a})\mskip1.5mu]} or \ensuremath{[\mskip1.5mu (\mathrm{2},\Varid{b})\mskip1.5mu]}, or even
\ensuremath{[\mskip1.5mu (\mathrm{2},\Varid{b}),(\mathrm{1},\Varid{a}),(\mathrm{2},\Varid{b})\mskip1.5mu]} are still acceptable outputs. So \ensuremath{\Varid{zip}}, as given in the Elixir documentation, is also
a relation.

A similar level of ambiguity is found in Haskell's
\emph{Data.List} library:
\begin{quote}\em
	Takes two lists and returns a list of corresponding pairs. If one input
	list is shorter than the other, excess elements of the longer list are discarded,
	even if one of the lists is infinite. \cite{base-4.19.1.0}
\end{quote}
Note how indefinite article ``a'' carries with it a relational interpretation.
However, the reference to ``excess elements'' is indicative of some concern
with extra positional information concerning the output.

The Javascript description is much more specific, which is noticeable in the use of definite articles:
\begin{quote}\em
	The zip method merges together the values of the given array with the values
	of the original collection at the corresponding index.
  \cite{collect-js}
\end{quote}
However, such ``the''s are problematic wherever the indices in one collection cannot be found in the other.
So, the meaning of \ensuremath{\Varid{zip}} in Javascript is that of a \emph{partial} function.
Finally, Python offers \ensuremath{\Varid{zip}} as follows:
\begin{quote}\em
	Make an iterator that aggregates elements from each of the iterables.  Returns
	an iterator of tuples, where the i-th tuple contains the i-th element from
	each of the argument sequences or iterables. The iterator stops when the
	shortest input iterable is exhausted. With a single iterable argument,
	it returns an iterator of 1-tuples. With no arguments, it returns an empty iterator.
  \cite{docs-python3.3-functions}
\end{quote}

Compared to the previous descritptions, Python's sounds over-specific in the
sense that it almost describes the actual implementation in natural language. Isn't
there some half-way informal description of \ensuremath{\Varid{zip}} that is unambiguous but
not such over-specific?
The answer may lie within Carroll Morgan's dictum, ``go formal informally'' \cite{Mo24},
which encapsulates the essence of achieving precision even in textual specifications,
emphasizing disciplined articulation without sacrificing readability.
How does one achieve such conciseness?

\section{Go formal informally} \label{sec:240531f}

Should the exercise above be repeated for the \ensuremath{\Varid{takeWhile}} function or method,
one would find textual specifications of similarly variable quality.
In particular, we would see that its name in Swift is \ensuremath{\Varid{prefix}} and that its description
in Haskell's Data.List package reads as follows:
\begin{eqnarray}
\begin{minipage}{0.85\linewidth}\em
	Applied to a predicate \ensuremath{\Varid{p}} and a list \ensuremath{\Varid{xs}}, returns the longest prefix (possibly
	empty) of \ensuremath{\Varid{xs}} of elements that satisfy \ensuremath{\Varid{p}}. \cite{base-4.19.1.0}
\end{minipage}
	\label{eq:240527a}
\end{eqnarray}

What catches the eye in this excerpt is its use of the \emph{concept}
of a \emph{list prefix}, together with an \emph{ordering} among prefixes that
brings with it the superlative ``longest''.
Note how clear the prose is: (a) \ensuremath{\Varid{takeWhile}\;\Varid{p}\;\Varid{xs}} is \emph{a}  prefix of
\ensuremath{\Varid{xs}}; (b) all \ensuremath{\Varid{x}} in such a prefix satisfy \ensuremath{\Varid{p}}; (c) \ensuremath{\Varid{takeWhile}\;\Varid{p}\;\Varid{xs}} is \emph{the}
longest such prefix. Also note how the indefinite articles of clauses (a) and (b)
become determined in clause (c).\footnote{Of course, one needs to be sure
that such \emph{longest prefixes} always exist, but that part of the exercise
lives in another level of the discussion, as we shall see later.}

Let us go deeper into the analysis of these clauses. First, let \ensuremath{x_1 \;\mathbin{\preccurlyeq}\;x_2 }
denote the fact that list \ensuremath{x_1 } is a prefix of \ensuremath{x_2 }.  So clause (a) becomes
\begin{eqnarray*}
\ensuremath{\Varid{takeWhile}\;\Varid{p}\;\Varid{xs}\;\mathbin{\preccurlyeq}\;\Varid{xs}}
\end{eqnarray*}
Clause (b) easily leads to:
~ \(
\ensuremath{\rcbnaked{\forall}{\Varid{x}}{\Varid{x}\;{\in}\;\Varid{takeWhile}\;\Varid{p}\;\Varid{xs}}{\Varid{p}\;\Varid{x}= \Conid{True}}}
\),
which can be made much shorter by use of the \ensuremath{\Varid{all}\;\Varid{p}} function in Haskell:
\begin{eqnarray}
\ensuremath{\Varid{all}\;\Varid{p}\;\Varid{xs}\mathrel{=}\rcbnaked{\forall}{\Varid{x}}{\Varid{x}\;{\in}\;\Varid{xs}}{\Varid{p}\;\Varid{x}= \Conid{True}}}
	\label{eq:240528b}
\end{eqnarray}
We can put (a) + (b) together as follows:
\begin{eqnarray}
\ensuremath{\Varid{ys}\mathrel{=}\Varid{takeWhile}\;\Varid{p}\;\Varid{xs}}
	& \ensuremath{\Rightarrow } &
\ensuremath{\begin{lcbr}\Varid{ys}\;\mathbin{\preccurlyeq}\;\Varid{xs}\\\Varid{all}\;\Varid{p}\;\Varid{ys}\end{lcbr}}
	\label{eq:240527b}
\end{eqnarray}

Now look at what clause (c) dictates: among all prefixes \ensuremath{\Varid{ys}} satisfying the
clauses on the right side of (\ref{eq:240527b}), \ensuremath{\Varid{ys}} is a prefix of \ensuremath{\Varid{takeWhile}\;\Varid{p}\;\Varid{xs}}, which is
the longest such prefix. Thus the converse (\ensuremath{\Leftarrow}) implication holds when replacing \ensuremath{\Varid{ys}\mathrel{=}\Varid{takeWhile}\;\Varid{p}\;\Varid{xs}} by
\ensuremath{\Varid{ys}\;\mathbin{\preccurlyeq}\;\Varid{takeWhile}\;\Varid{p}\;\Varid{xs}}, which is clause (c) put formally in symbols.
This leads to the following formal specification of the (informal) Data.List
description of \ensuremath{\Varid{takeWhile}} given in the Hackage repository:
\begin{eqnarray}
\ensuremath{\begin{lcbr}\Varid{ys}\;\mathbin{\preccurlyeq}\;\Varid{xs}\\\Varid{all}\;\Varid{p}\;\Varid{ys}\end{lcbr}}
        \ensuremath{\:\Leftrightarrow\:}
\ensuremath{\Varid{ys}\;\mathbin{\preccurlyeq}\;\Varid{takeWhile}\;\Varid{p}\;\Varid{xs}}
	\label{eq:240527c}
\end{eqnarray}
We conclude that the clarity of the prose in (\ref{eq:240527a}) is a good example of
``go formal informally'' \cite{Mo24}.

As it turns out, (\ref{eq:240527a}) and (\ref{eq:240527c}) match a common
\emph{specification pattern} in which the description of a programming problem
is divided into an \emph{easy part} --- clauses (a) + (b) above --- which offers
many solutions (a binary relation), out of which the best solution is extracted
--- the \emph{hard} part \cite{MO12a}. In the \ensuremath{\Varid{takeWhile}} example above
the hard part is given by clause (c).

It is interesting to compare (\ref{eq:240527a}) with the descriptions of the
same function in the libraries of other languages:

\begin{itemize}
\item Javascript: 
\begin{quote}\em
The \ensuremath{\Varid{takeWhile}} method returns items in the collection until the given callback returns false.
\cite{collect-js}
\end{quote}
\item Python:
\begin{quote}\em
Make an iterator that returns elements from the iterable as long as the predicate
is true.
\cite{docs-python3.3-itertools}
\end{quote}
\item Swift:
\begin{quote}\em
Returns a subsequence containing the initial elements until predicate returns false and
skipping the remaining elements\footnote{As already mentioned, Swift's \ensuremath{\Varid{takeWhile}} function is called \ensuremath{\Varid{prefix}}.}. \cite{apple-swift}
\end{quote}
\item Elixir:
\begin{quote}\em
\ensuremath{take\_while\;(\Varid{enumerable},\Varid{fun})}
Takes the elements from the beginning of the \ensuremath{\Varid{enumerable}} while \ensuremath{\Varid{fun}} returns a truthy value. \cite{elixir/1.12}
\end{quote}
\item F\#:
\begin{quote}\em
[To use] Seq.takeWhile specify a predicate function (a Boolean function) and create a sequence 
from another sequence made up of those elements of the original sequence for which the predicate is true, 
but stop before the first element for which the predicate returns false.
\cite{ms-f-sharp}
\end{quote}
\end{itemize}

Clearly, Javascript, Python and F\# do not ensure \ensuremath{\Varid{ys}\;\mathbin{\preccurlyeq}\;\Varid{xs}} in (\ref{eq:240527c});
Swift uses the \emph{subsequence} relation, which is more general than prefix, and therefore needs to
add ``containing the initial elements'' to achieve the required precison; finally, Elixir's description
is more algorithmic, but ``taking the elements from the beginning'' captures the prefix concept.

It turns out that the specification pattern found in \ensuremath{\Varid{takeWhile}}
(\ref{eq:240527a},\ref{eq:240527c}) is an instance of a \emph{Galois connection},
a mathematical device due to the French matematician Evariste Galois
(1811-1832). References \cite{MO12a,Ol23} give more instances of this pattern and develop
the corresponding theory.

In this paper we are more interested in sending a broader message to the programming
community, in particular to functionl programmers, about the need for precision
in library documentation and its link to formal correctness. Starting from
the \ensuremath{\Varid{takeWhile}} example, the rest of this paper will show how productive
such a pattern is, searching for instances of it in Haskell libraries.
Without further ado, let "Haskell meet Evariste" and learn from what the latter has to offer.

\section{\ensuremath{\Varid{takeWhile}} continued} \label{sec:takeWhile} 

In the previous section, (\ref{eq:240527a}) --- resp.\ (\ref{eq:240527c}) --- was regarded as an
informal --- resp.\ formal --- description of the meaning of function (method) \ensuremath{\Varid{takeWhile}},
which can be regarded as interfaces needed for programmers to decide about reusing the function in the programs they are writing.
This understanding of \emph{what the function does} is related to its
properties. Indeed, many entries in the Haskell libraries resort to
properties to better complement the given textual descriptions,
see for instance \ensuremath{\Varid{foldr}} and \ensuremath{\Varid{foldl}} \cite{base-4.19.1.0}.

We claim that a specification such as (\ref{eq:240527c}) --- and its rendering
in precise prose (\ref{eq:240527a}) --- plays the role of a \emph{universal
property} wherefrom other properties can be extracted. Furthermore, implementations
can also be derived from such universal properties, as shown in \cite{Ol23,PP23}.
Indeed, later on we will show that the standard implementation
\begin{quote}
\begin{hscode}\SaveRestoreHook
\column{B}{@{}>{\hspre}l<{\hspost}@{}}%
\column{13}{@{}>{\hspre}l<{\hspost}@{}}%
\column{25}{@{}>{\hspre}l<{\hspost}@{}}%
\column{28}{@{}>{\hspre}l<{\hspost}@{}}%
\column{E}{@{}>{\hspre}l<{\hspost}@{}}%
\>[B]{}\Varid{takeWhile}{}\<[25]%
\>[25]{}\mathbin{::}(\Varid{a}{\,\rightarrow\,}\Conid{Bool}){\,\rightarrow\,}[\mskip1.5mu \Varid{a}\mskip1.5mu]{\,\rightarrow\,}[\mskip1.5mu \Varid{a}\mskip1.5mu]{}\<[E]%
\\
\>[B]{}\Varid{takeWhile}\;\anonymous \;[\mskip1.5mu \mskip1.5mu]{}\<[25]%
\>[25]{}\mathrel{=}{}\<[28]%
\>[28]{}[\mskip1.5mu \mskip1.5mu]{}\<[E]%
\\
\>[B]{}\Varid{takeWhile}\;\Varid{p}\;(\Varid{x}\mathbin{:}\Varid{xs}){}\<[E]%
\\
\>[B]{}\hsindent{13}{}\<[13]%
\>[13]{}\mid \Varid{p}\;\Varid{x}{}\<[25]%
\>[25]{}\mathrel{=}{}\<[28]%
\>[28]{}\Varid{x}\mathbin{:}\Varid{takeWhile}\;\Varid{p}\;\Varid{xs}{}\<[E]%
\\
\>[B]{}\hsindent{13}{}\<[13]%
\>[13]{}\mid \Varid{otherwise}\mathrel{=}{}\<[28]%
\>[28]{}[\mskip1.5mu \mskip1.5mu]{}\<[E]%
\ColumnHook
\end{hscode}\resethooks
\end{quote}
found in the Data.List package is derivable from
(\ref{eq:240527c}), meaning that implementation and formal specification
match formally. Put in other words, the code is correct with respect 
to its formal specification (\ref{eq:240527c}).

Prior to such a derivation, let us guess possible properties of \ensuremath{\Varid{takeWhile}}
that could enrich its Data.List description. For instance, what happens when
one chains two \ensuremath{\Varid{takeWhiles}}? One will expect
\(
	\ensuremath{\Varid{takeWhile}\;\Varid{p}\comp \Varid{takeWhile}\;\Varid{q}\mathrel{=}\Varid{takeWhile}\;(\Varid{p}\mathrel{\wedge}\Varid{q})}
\) to hold,
where \ensuremath{(\Varid{p}\mathrel{\wedge}\Varid{q})\;\Varid{x}\mathrel{=}(\Varid{p}\;\Varid{x})\mathrel{\wedge}(\Varid{q}\;\Varid{x})} --- note the harmless overloading of \ensuremath{(\mathrel{\wedge})}.
The same guess in pointwise Haskell goes like this:
\begin{eqnarray}
	\ensuremath{\Varid{takeWhile}\;\Varid{p}\;(\Varid{takeWhile}\;\Varid{q}\;\Varid{xs})\mathrel{=}\Varid{takeWhile}\;(\Varid{p}\mathrel{\wedge}\Varid{q})\;\Varid{xs}}
	\label{eq:240528a}
\end{eqnarray}

To approve or reject the above assumption, one can try starting with (\ref{eq:240527a})
but difficulties in the nesting of superlatives (``longest of the longest'', and so on)
will get in the way. Starting at (\ref{eq:240527c}) will overcome such
obstacles, but we need to know some extra formalities regarding list prefixes.

Let us accept\footnote{Here we mean \emph{informally accept}, i.e.\ with
no formal evidence given in this paper. For such formal evidences please
consult \cite{PP23}.} that prefix is a partial order, that is:
(a) any list is a prefix of itself, i.e.\ \ensuremath{\Varid{x}\;\mathbin{\preccurlyeq}\;\Varid{x}} always holds:
(b) a prefix of a prefix is a prefix (transitivity of the concept);
(c) ``symmetric prefixes'' mean equality: \ensuremath{\Varid{x}\;\mathbin{\preccurlyeq}\;\Varid{y}} and \ensuremath{\Varid{y}\;\mathbin{\preccurlyeq}\;\Varid{x}}
entail \ensuremath{\Varid{x}\mathrel{=}\Varid{y}} (antisymmetry).

Now imagine that you have two lists \ensuremath{\Varid{xs}} and \ensuremath{\Varid{ys}} that have \emph{exactly
the same} prefixes. That is, if some \ensuremath{\Varid{zs}} is a prefix of \ensuremath{\Varid{xs}}, than it is
a prefix of \ensuremath{\Varid{ys}}, and vice-versa. Intuitively, \ensuremath{\Varid{xs}} and \ensuremath{\Varid{ys}} will have to be the same
list, and indeed they are. This is known as the \emph{indirect equality}
principle, which can already be found in Dijkstra's technical notes \cite{Di01},
its use in maths being of course much older.\footnote{In classical set-theory,
for instance, one says that two sets \ensuremath{\Conid{A}} and \ensuremath{\Conid{B}} are the same if they have
exactly the same elements, that is, \ensuremath{\Varid{a}\;{\in}\;\Conid{A}} is equivalent to \ensuremath{\Varid{a}\;{\in}\;\Conid{B}},
for any \ensuremath{\Varid{a}}.} Put formally, this principle is as follows:\footnote{Note
that this principle is not a privilege of list prefixing: it works for \emph{any
partial order}.}
\begin{eqnarray}
	\ensuremath{\Varid{xs}\mathrel{=}\Varid{ys}}
& \ensuremath{\:\Leftrightarrow\:} &
	\ensuremath{\rcbnaked{\forall}{\Varid{zs}}{}{(\Varid{zs}\;\mathbin{\preccurlyeq}\;\Varid{xs}\:\Leftrightarrow\:\Varid{zs}\;\mathbin{\preccurlyeq}\;\Varid{ys})}}
	\label{eq:240528c}
\end{eqnarray}

As we shall briefly see, this is all we need to know to approve (\ref{eq:240528a}).
Note that we have no Haskell code for \ensuremath{\Varid{takeWhile}} yet, so attempting an inductive formal
proof in say Agda or Coq is ruled out.
All we know about \ensuremath{\Varid{takeWhile}}
is its occurrence on the right side of a prefix symbol in (\ref{eq:240527c}).
So we have two possible starting points: either
    \ensuremath{\Varid{zs}\;\mathbin{\preccurlyeq}\;\Varid{takeWhile}\;\Varid{p}\;(\Varid{takeWhile}\;\Varid{q}\;\Varid{xs})}
or
    \ensuremath{\Varid{zs}\;\mathbin{\preccurlyeq}\;\Varid{takeWhile}\;(\Varid{p}\;\land\;\Varid{q})\;\Varid{xs}}.
Let us choose the first option:
\begin{eqnarray*}
\start
    \ensuremath{\Varid{zs}\;\mathbin{\preccurlyeq}\;\Varid{takeWhile}\;\Varid{p}\;(\Varid{takeWhile}\;\Varid{q}\;\Varid{xs})}
\just\equiv{ (\ref{eq:240527c}) concerning the outermost \ensuremath{\Varid{takeWhile}} }
    \ensuremath{\Varid{all}\;\Varid{p}\;\Varid{zs}\;\land\;\Varid{zs}\;\mathbin{\preccurlyeq}\;\Varid{takeWhile}\;\Varid{q}\;\Varid{xs}}
\just\equiv{ (\ref{eq:240527c}) concerning the other \ensuremath{\Varid{takeWhile}} }
    \ensuremath{\Varid{all}\;\Varid{p}\;\Varid{zs}\;\land\;(\Varid{all}\;\Varid{q}\;\Varid{zs}\;\land\;\Varid{zs}\;\mathbin{\preccurlyeq}\;\Varid{xs})}
\just\equiv{ associativity of conjunction }
    \ensuremath{(\Varid{all}\;\Varid{p}\;\Varid{zs}\;\land\;\Varid{all}\;\Varid{q}\;\Varid{zs})\;\land\;\Varid{zs}\;\mathbin{\preccurlyeq}\;\Varid{xs}}
	\eqnnewpage
\just\equiv{ property of universal quantification, cf (\ref{eq:240528b}) }
    \ensuremath{(\Varid{all}\;(\Varid{p}\;\land\;\Varid{q})\;\Varid{zs})\;\land\;\Varid{zs}\;\mathbin{\preccurlyeq}\;\Varid{xs}}
\just\equiv{ (\ref{eq:240527c}) again }
    \ensuremath{\Varid{zs}\;\mathbin{\preccurlyeq}\;\Varid{takeWhile}\;(\Varid{p}\;\land\;\Varid{q})\;\Varid{xs}}
%
\end{eqnarray*}
Noting that \ensuremath{\Varid{zs}} is universally quantified above, the equivalence of the
starting and ending points of the reasoning instantiates the righthand clause 
of (\ref{eq:240528c}). Taking the left part of the same clause, we immediately
approve the equality guessed in (\ref{eq:240528a}).

Another property that one might wish to add to the \ensuremath{\Varid{takeWhile}} entry in Data.List
is \ensuremath{\Varid{takeWhile}\;\anonymous \;[\mskip1.5mu \mskip1.5mu]\mathrel{=}[\mskip1.5mu \mskip1.5mu]}. For this (\ref{eq:240527a}) would be ``enough'', since the only
prefix of the empty list is itself.
But (\ref{eq:240527c}) gives us even simpler evidence:
\begin{eqnarray*}
\start
\ensuremath{\Varid{ys}\;\mathbin{\preccurlyeq}\;\Varid{takeWhile}\;\anonymous \;[\mskip1.5mu \mskip1.5mu]}
\just\equiv{ again (\ref{eq:240527c}) }
\ensuremath{\Varid{ys}\;\mathbin{\preccurlyeq}\;[\mskip1.5mu \mskip1.5mu]\mathrel{\wedge}\Varid{all}\;\anonymous \;[\mskip1.5mu \mskip1.5mu]}
\just\equiv{ \ensuremath{\Varid{all}\;\anonymous \;[\mskip1.5mu \mskip1.5mu]\mathrel{=}\Conid{True}} (empty range in universal quantifier) }
\ensuremath{\Varid{ys}\;\mathbin{\preccurlyeq}\;[\mskip1.5mu \mskip1.5mu]}
\just{::}{ indirect equality (\ref{eq:240528c}) }
\ensuremath{[\mskip1.5mu \mskip1.5mu]\mathrel{=}\Varid{takeWhile}\;\anonymous \;[\mskip1.5mu \mskip1.5mu]}
\end{eqnarray*}

Note that \ensuremath{\Varid{takeWhile}\;\anonymous \;[\mskip1.5mu \mskip1.5mu]\mathrel{=}[\mskip1.5mu \mskip1.5mu]} can be regarded as the base case of the Haskell
code we now wish to derive from (\ref{eq:240527c}). By pattern matching we
are left with filling the dots in the other case, cf.
\begin{quote}
\begin{hscode}\SaveRestoreHook
\column{B}{@{}>{\hspre}l<{\hspost}@{}}%
\column{E}{@{}>{\hspre}l<{\hspost}@{}}%
\>[B]{}\Varid{takeWhile}\;\Varid{p}\;(\Varid{x}\mathbin{:}\Varid{xs})\mathrel{=}\mathbin{......}{}\<[E]%
\ColumnHook
\end{hscode}\resethooks
\end{quote}
By the same procedure as above, we have to start from \ensuremath{\Varid{ys}\;\mathbin{\preccurlyeq}\;\Varid{takeWhile}\;\Varid{p}\;(\Varid{x}\mathbin{:}\Varid{xs})} and see what happens.
But we need something else: what can we say about \ensuremath{\Varid{ys}\;\mathbin{\preccurlyeq}\;(\Varid{x}\mathbin{:}\Varid{xs})}?

Note that \ensuremath{[\mskip1.5mu \mskip1.5mu]\;\mathbin{\preccurlyeq}\;\Varid{xs}} always holds, meaning that the empty list is the least element of the prefix ordering.
So, either \ensuremath{\Varid{ys}\mathrel{=}[\mskip1.5mu \mskip1.5mu]} and we are done, or \ensuremath{\Varid{ys}} should start with \ensuremath{\Varid{x}} at the front and its tail be a prefix of
\ensuremath{\Varid{xs}}. (See \cite{Ol23} for a formal justification of these clauses.) In symbols, we have:
\begin{eqnarray}
	\ensuremath{\Varid{ys}\;\mathbin{\preccurlyeq}\;(\Varid{x}\mathbin{:}\Varid{xs})\:\Leftrightarrow\:\Varid{ys}\mathrel{=}[\mskip1.5mu \mskip1.5mu]\;\lor\;\langle\, \exists\:\Varid{zs}\: : \:\Varid{ys}\mathrel{=}(\Varid{x}\mathbin{:}\Varid{zs})\: : \:\Varid{zs}\;\mathbin{\preccurlyeq}\;\Varid{xs}\,\rangle}
	\label{eq:pw-prefix} 
\end{eqnarray}
Armed with (\ref{eq:pw-prefix}) we proceed as before:
\begin{eqnarray*}
\start
     \ensuremath{\Varid{ys}\;\mathbin{\preccurlyeq}\;\Varid{takeWhile}\;\Varid{p}\;(\Varid{x}\mathbin{:}\Varid{xs})}
\just\equiv{ again (\ref{eq:240527c}) }
    \ensuremath{\Varid{all}\;\Varid{p}\;\Varid{ys}\;\land\;\Varid{ys}\;\mathbin{\preccurlyeq}\;(\Varid{x}\mathbin{:}\Varid{xs})}
\just\equiv{ (\ref{eq:pw-prefix}) above }
	\eqnnewpage
    \ensuremath{\Varid{all}\;\Varid{p}\;\Varid{ys}\;\land\;(\Varid{ys}\mathrel{=}[\mskip1.5mu \mskip1.5mu]\;\lor\;\langle\, \exists\:\Varid{zs}\: : \:\Varid{ys}\mathrel{=}(\Varid{x}\mathbin{:}\Varid{zs})\: : \:\Varid{zs}\;\mathbin{\preccurlyeq}\;\Varid{xs}\,\rangle)}
\just\equiv{ \ensuremath{\mathrel{\wedge}}/\ensuremath{\lor} distribution }
    \ensuremath{(\Varid{all}\;\Varid{p}\;\Varid{ys}\;\land\;\Varid{ys}\mathrel{=}[\mskip1.5mu \mskip1.5mu])\;\lor\;\langle\, \exists\:\Varid{zs}\: : \:\Varid{ys}\mathrel{=}(\Varid{x}\mathbin{:}\Varid{zs})\: : \:\Varid{all}\;\Varid{p}\;\Varid{ys}\;\land\;(\Varid{zs}\;\mathbin{\preccurlyeq}\;\Varid{xs})\,\rangle}
\just\equiv{ \ensuremath{\Varid{all}\;\Varid{p}\;\Varid{ys}\,\Leftarrow\,\Varid{ys}\mathrel{=}[\mskip1.5mu \mskip1.5mu]} }
    \ensuremath{\Varid{ys}\mathrel{=}[\mskip1.5mu \mskip1.5mu]\;\lor\;\langle\, \exists\:\Varid{zs}\: : \:\Varid{ys}\mathrel{=}(\Varid{x}\mathbin{:}\Varid{zs})\: : \:\Varid{all}\;\Varid{p}\;\Varid{ys}\;\land\;(\Varid{zs}\;\mathbin{\preccurlyeq}\;\Varid{xs})\,\rangle}
\just\equiv{ \ensuremath{\Varid{all}\;\Varid{p}\;(\Varid{x}\mathbin{:}\Varid{xs})\;\,\equiv\,\;\Varid{p}\;\Varid{x}\;\land\;\Varid{all}\;\Varid{p}\;\Varid{xs}} }
    \ensuremath{\Varid{ys}\mathrel{=}[\mskip1.5mu \mskip1.5mu]\;\lor\;\langle\, \exists\:\Varid{zs}\: : \:\Varid{ys}\mathrel{=}(\Varid{x}\mathbin{:}\Varid{zs})\: : \:\Varid{p}\;\Varid{x}\;\land\;\Varid{all}\;\Varid{p}\;\Varid{zs}\;\land\;\Varid{zs}\;\mathbin{\preccurlyeq}\;\Varid{xs}\,\rangle}
\just\equiv{ (\ref{eq:240527c}) once again }
    \ensuremath{\Varid{ys}\mathrel{=}[\mskip1.5mu \mskip1.5mu]\;\lor\;\langle\, \exists\:\Varid{zs}\: : \:\Varid{ys}\mathrel{=}(\Varid{x}\mathbin{:}\Varid{zs})\: : \:\Varid{p}\;\Varid{x}\;\land\;\Varid{zs}\;\mathbin{\preccurlyeq}\;\Varid{takeWhile}\;\Varid{p}\;\Varid{xs}\,\rangle}
%
\end{eqnarray*}
\begin{itemize}
\item case \ensuremath{\Varid{p}\;\Varid{x}} holds:
\begin{eqnarray*}
\start
    \ensuremath{\Varid{zs}\mathrel{=}[\mskip1.5mu \mskip1.5mu]\;\lor\;\langle\, \exists\:\Varid{zs}\: : \:\Varid{zs}\mathrel{=}(\Varid{x}\mathbin{:}\Varid{zs})\: : \:\Varid{zs}\;\mathbin{\preccurlyeq}\;\Varid{takeWhile}\;\Varid{p}\;\Varid{xs}\,\rangle}
\just\equiv{ (\ref{eq:pw-prefix}) }
    \ensuremath{\Varid{zs}\;\mathbin{\preccurlyeq}\;\Varid{x}\mathbin{:}\Varid{takeWhile}\;\Varid{p}\;\Varid{xs}}
\just{::}{ indirect equality (\ref{eq:240528c}) }
    \ensuremath{\Varid{takeWhile}\;\Varid{p}\;(\Varid{x}\mathbin{:}\Varid{xs})\mathrel{=}\Varid{x}\mathbin{:}\Varid{takeWhile}\;\Varid{p}\;\Varid{xs}}
\end{eqnarray*}
\item case \ensuremath{\neg \;(\Varid{p}\;\Varid{x})}:
\begin{eqnarray*}
\start
    \ensuremath{\Varid{zs}\mathrel{=}[\mskip1.5mu \mskip1.5mu]}
\just\equiv{ by antisymmetry,  since \ensuremath{[\mskip1.5mu \mskip1.5mu]\;\mathbin{\preccurlyeq}\;\Varid{zs}} always holds }
    \ensuremath{\Varid{zs}\;\mathbin{\preccurlyeq}\;[\mskip1.5mu \mskip1.5mu]}
\just{::}{ indirect equality (\ref{eq:240528c}) }
    \ensuremath{\Varid{takeWhile}\;\Varid{p}\;(\Varid{x}\mathbin{:}\Varid{t})\mathrel{=}[\mskip1.5mu \mskip1.5mu]}
\end{eqnarray*}
\end{itemize}
Thus one gets, from (\ref{eq:240527c}), the Haskell code that is given in Data.List. 
Note that this code is ``correct-by-construction'' with respect to its specification.

\section{Generalization}
The \ensuremath{\Varid{takeWhile}} case-study of the previous sections shows how a ``good start''
in describing the meaning of a piece of code pays off. Clear-cut, well understood
\emph{concepts} \cite{Ja21} play a major role. (In the \ensuremath{\Varid{takeWhile}} case,
the concept of a list \emph{prefix} is what made the description more effective
and more rigorous than the others.) Unfortunately, most descriptions in software
libraries are too vague to yield the same outcome.

Clumsy or vague prose lead to erroneous understanding, let alone
inadequate or convoluted formal specifications. It is the experience of
many that handling such complex specifications may not pay the effort, leading
to a certain reluctance towards them. Clearly, there exists a gap in program
documentation --- on the one hand, some exhibit insufficient rigor, while on
the other hand, others manifest a level of rigor that many choose to deem
excessively complex. This is why expressing problem descriptions in the
\emph{easy-hard-split} way proposed by \cite{MO12a} is useful, because it
has a simple structure and a rich (generic) theory that can be re-used in different contexts.

Still in the list-processing area, we observe that \ensuremath{\Varid{take}} in Data.List is also
described using the prefix concept:
\begin{quote}\em
	\ensuremath{\Varid{take}\;\Varid{n}}, applied to a list \ensuremath{\Varid{xs}}, returns the prefix of \ensuremath{\Varid{xs}} of length \ensuremath{\Varid{n}}, 
        or \ensuremath{\Varid{xs}} itself if \ensuremath{\Varid{n}\geq \mathsf{length}\;\Varid{xs}}. \cite{base-4.19.1.0}
\end{quote}
In this case, however, the prose is of lesser quality and, as shown in \cite{Ol23}, it might be written otherwise:
\begin{quote}\em
\ensuremath{\Varid{take}\;\Varid{n}\;\Varid{xs}} should yield the {longest} {prefix} of \ensuremath{\Varid{xs}} not exceeding \ensuremath{\Varid{n}} in {length}.
\end{quote}
From this alternative text one immediately writes the corresponding {formal description},
\begin{eqnarray*}
	\ensuremath{\underbrace{\mathsf{length}\;\Varid{ys}\leq\Varid{n}\mathrel{\wedge}\Varid{ys}\;\mathbin{\preccurlyeq}\;\Varid{xs}}_{\Varid{easy}}} & \equiv & \ensuremath{\underbrace{\Varid{ys}\;\mathbin{\preccurlyeq}\;\Varid{take}\;\Varid{n}\;\Varid{xs}}_{\Varid{hard}}}
\end{eqnarray*}
which --- following the same \emph{easy-hard-split} pattern of (\ref{eq:240527c})
--- has similar yield, as shown in \cite{Ol23}. 

Back to the \ensuremath{\Varid{zip}} function, so poorly described in most libraries as shown in section \ref{sec:motivation}, we go the
other way around --- this time we first write a formal description, again relying on the prefix concept:
\begin{equation}\label{eq:gc-zip}
	\ensuremath{\begin{lcbr}\mathsf{map}\;\mathit{fst}\;\Varid{zs}\;\mathbin{\preccurlyeq}\;\Varid{xs}\\\mathsf{map}\;\mathit{snd}\;\Varid{zs}\;\mathbin{\preccurlyeq}\;\Varid{ys}\end{lcbr}\;\,\equiv\,\;\Varid{zs}\;\mathbin{\preccurlyeq}\;\Varid{zip}\;\Varid{xs}\;\Varid{ys}}
\end{equation}
Then we express it in words:
\begin{quote}\em
\ensuremath{\Varid{zip}\;\Varid{xs}\;\Varid{ys}} is the longest possible list of pairs that yields prefixes of both \ensuremath{\Varid{xs}} and
\ensuremath{\Varid{ys}} once its pairs are projected by \ensuremath{\mathit{fst}} and \ensuremath{\mathit{snd}}, respectively.
\end{quote}
Knowing
\begin{eqnarray}
	\ensuremath{\Varid{unzip}\;\Varid{zs}\mathrel{=}(\mathsf{map}\;\mathit{fst}\;\Varid{zs},\mathsf{map}\;\mathit{snd}\;\Varid{zs})}
	\label{eq:260212a}
\end{eqnarray}
one can express (\ref{eq:gc-zip}) as follows,
\begin{eqnarray*}
	\ensuremath{\mathbf{let}\;(\Varid{x},\Varid{y})\mathrel{=}\Varid{unzip}\;\Varid{zs}\;\mathbf{in}\;\begin{lcbr}\Varid{x}\;\mathbin{\preccurlyeq}\;\Varid{xs}\\\Varid{y}\;\mathbin{\preccurlyeq}\;\Varid{ys}\end{lcbr}\;\,\equiv\,\;\Varid{zs}\;\mathbin{\preccurlyeq}\;\Varid{zip}\;\Varid{xs}\;\Varid{ys}}
\end{eqnarray*}
making it apparent that \ensuremath{\Varid{unzip}} is a kind of inverse of \ensuremath{\Varid{zip}}. But only a right inverse,
\begin{eqnarray}
\ensuremath{\Varid{zip}\;\Varid{xs}\;\Varid{ys}\mathrel{=}\Varid{zs}\;\mathbf{where}\;(\Varid{xs},\Varid{ys})\mathrel{=}\Varid{unzip}\;\Varid{zs}}
	\label{eq:240528d}
\end{eqnarray}
as the converse property is easily dismissed everytime one zips lists of different lengths.

How can one be sure that (\ref{eq:240528d}) holds without access to the Haskell code
of \ensuremath{\Varid{zip}}? Thanks to (\ref{eq:260212a}) we can do the corresponding substitutions
in (\ref{eq:240528d}) to obtain:
\begin{eqnarray}
\ensuremath{\Varid{zip}\;(\mathsf{map}\;\mathit{fst}\;\Varid{zs})\;(\mathsf{map}\;\mathit{snd}\;\Varid{zs})\mathrel{=}\Varid{zs}}
	\label{eq:240531d}
\end{eqnarray}
Let us plug this into (\ref{eq:gc-zip}) and see what happens:
\begin{eqnarray*}
\start
\ensuremath{\Varid{zs}\;\mathbin{\preccurlyeq}\;\Varid{zip}\;(\mathsf{map}\;\mathit{fst}\;\Varid{zs})\;(\mathsf{map}\;\mathit{snd}\;\Varid{zs})}
\just\equiv{ (\ref{eq:gc-zip}) }
\ensuremath{\begin{lcbr}\mathsf{map}\;\mathit{fst}\;\Varid{zs}\;\mathbin{\preccurlyeq}\;\mathsf{map}\;\mathit{fst}\;\Varid{zs}\\\mathsf{map}\;\mathit{snd}\;\Varid{zs}\;\mathbin{\preccurlyeq}\;\mathsf{map}\;\mathit{snd}\;\Varid{zs}\end{lcbr}}
\just\equiv{ \ensuremath{\Varid{x}\;\mathbin{\preccurlyeq}\;\Varid{x}} holds  for \textbf{every} \ensuremath{\Varid{x}} }
\ensuremath{\Conid{True}}
\end{eqnarray*}

So we get \ensuremath{\Varid{zs}\;\mathbin{\preccurlyeq}\;\Varid{zip}\;(\mathsf{map}\;\mathit{fst}\;\Varid{zs})\;(\mathsf{map}\;\mathit{snd}\;\Varid{zs})}, which is not yet (\ref{eq:240531d}) but is not far from it.
(See section \ref{sec:240531e} for how to actually reach (\ref{eq:240531d}).) 

Note that, again, we are deriving properties of a function, \ensuremath{\Varid{zip}} in this
case, prior to encoding it or knowing its Haskell code. The calculation above
is one of the standard ways to infer properties from Galois connections known
as \emph{cancellation} \cite{Ba04a}. In essence, we made
the substitutions that render the left side of (\ref{eq:gc-zip}) trivially
true (via the reflexivity of the prefix ordering), i.e.\ we \emph{cancelled}
it. This is named \emph{left-cancellation} because it worked on the left side
of the equivalence.

Just to see what the dual \emph{right}-cancellation amounts to, let us just make
\ensuremath{\Varid{zs}\mathbin{:=}\Varid{zip}\;\Varid{xs}\;\Varid{ys}} in (\ref{eq:gc-zip}) to obtain:
\begin{eqnarray*}
\ensuremath{\begin{lcbr}\mathsf{map}\;\mathit{fst}\;(\Varid{zip}\;\Varid{xs}\;\Varid{ys})\;\mathbin{\preccurlyeq}\;\Varid{xs}\\\mathsf{map}\;\mathit{snd}\;(\Varid{zip}\;\Varid{xs}\;\Varid{ys})\;\mathbin{\preccurlyeq}\;\Varid{ys}\end{lcbr}\;\,\equiv\,\;\Varid{zip}\;\Varid{xs}\;\Varid{ys}\;\mathbin{\preccurlyeq}\;\Varid{zip}\;\Varid{xs}\;\Varid{ys}}
\end{eqnarray*}
Now the righthand side is trivially true and goes away and one is left with:
\begin{eqnarray*}
\ensuremath{\begin{lcbr}\mathsf{map}\;\mathit{fst}\;(\Varid{zip}\;\Varid{xs}\;\Varid{ys})\;\mathbin{\preccurlyeq}\;\Varid{xs}\\\mathsf{map}\;\mathit{snd}\;(\Varid{zip}\;\Varid{xs}\;\Varid{ys})\;\mathbin{\preccurlyeq}\;\Varid{ys}\end{lcbr}}
\end{eqnarray*}
This just says that \ensuremath{\Varid{zip}\;\Varid{xs}\;\Varid{ys}} is a ``solution'' to the easy part of the description:
if you project it by \ensuremath{\mathit{fst}} or \ensuremath{\mathit{snd}} you get prefixes of the lists that you have zipped.

\section{Changing the ordering}

Several other examples like the previous ones were taken from Hackage's Data.List
package and worked out in full detail
in a recent master thesis \cite{PP23}, trying to see how wide the scope of
the \emph{easy-hard-split} pattern of \cite{MO12a} is, supported by Galois-connection
theory. (Many of the technical details intentionally omitted from the current
paper to reach a wider audience can be found there.) The work looked at each
case along three layers of increasing formality: textual description, formal
\emph{pointwise} description (as above) and formal \emph{pointfree} description
using relation algebra \cite{BM97,Ba04a,MO12a}.

One of the (to be expected) conclusions is that the list {prefix} ordering is only applicable to functions
which wish to preserve contiguous elements from the start of the input lists. A function such as \ensuremath{\Varid{filter}}, described
as follows in Data.List,
\begin{eqnarray}
\begin{minipage}{0.85\linewidth}\em
Applied to a predicate and a list, returns the list of those elements that satisfy the predicate
\cite{base-4.19.1.0}
\end{minipage}
	\label{eq:240529a}
\end{eqnarray}
will not yield a prefix of the input list in most cases. 
For instance, \ensuremath{\Varid{filter}\;\Varid{even}\;[\mskip1.5mu \mathrm{2},\mathrm{4},\mathrm{5}\mskip1.5mu]\mathrel{=}[\mskip1.5mu \mathrm{2},\mathrm{4}\mskip1.5mu]} is a prefix of  
\ensuremath{[\mskip1.5mu \mathrm{2},\mathrm{4},\mathrm{5}\mskip1.5mu]} but  \ensuremath{\Varid{filter}\;\Varid{odd}\;[\mskip1.5mu \mathrm{2},\mathrm{4},\mathrm{5}\mskip1.5mu]\mathrel{=}[\mskip1.5mu \mathrm{5}\mskip1.5mu]} is not.

Does this mean that \ensuremath{\Varid{filter}} cannot be framed into the \emph{easy-hard-split} 
description pattern that has been used so far? 
No, it does fit in. What is required is a different \emph{concept} ---
that of a \emph{sublist} instead of \emph{prefix}. For instance, \ensuremath{[\mskip1.5mu \mathrm{5}\mskip1.5mu]} is not a prefix of \ensuremath{[\mskip1.5mu \mathrm{2},\mathrm{4},\mathrm{5}\mskip1.5mu]} 
but it is one of its sublists. Interestingly, \ensuremath{\Varid{takeWhile}} in (\ref{eq:240527c}) "becomes" \ensuremath{\Varid{filter}}
once prefixes give room to sublists,
\begin{eqnarray}
\ensuremath{\begin{lcbr}\Varid{ys}\;\mathbin{\sqsubseteq}\;\Varid{xs}\\\Varid{all}\;\Varid{p}\;\Varid{ys}\end{lcbr}}
        \ensuremath{\:\Leftrightarrow\:}
\ensuremath{\Varid{ys}\;\mathbin{\sqsubseteq}\;\Varid{filter}\;\Varid{p}\;\Varid{xs}}
	\label{eq:240530a}
\end{eqnarray}
where \ensuremath{\Varid{y}\;\mathbin{\sqsubseteq}\;\Varid{x}} means "\ensuremath{\Varid{y}} is a sublist of \ensuremath{\Varid{x}}".

Thus, a better description of \ensuremath{\Varid{filter}} in the Haskell's Data.List package 
could be just that of \ensuremath{\Varid{takeWhile}} (\ref{eq:240527a}) with "prefix" replaced by "sublist":
\begin{eqnarray*}
\begin{minipage}{0.85\linewidth}\em
	Applied to a predicate \ensuremath{\Varid{p}} and a list \ensuremath{\Varid{xs}}, returns the longest sublist (possibly
	empty) of \ensuremath{\Varid{xs}} of elements that satisfy \ensuremath{\Varid{p}}.
\end{minipage}
\end{eqnarray*}
Now, what makes a sublist different from a prefix?
For all \ensuremath{\Varid{xs}}, \ensuremath{[\mskip1.5mu \mskip1.5mu]\;\mathbin{\sqsubseteq}\;\Varid{xs}} also holds, meaning that the empty list is
again the least element of the ordering. For \ensuremath{\Varid{ys}\;\mathbin{\sqsubseteq}\;(\Varid{x}\mathbin{:}\Varid{xs})} we look
back at (\ref{eq:pw-prefix}) and do a little change:
\begin{eqnarray}
	\ensuremath{\Varid{ys}\;\mathbin{\sqsubseteq}\;(\Varid{x}\mathbin{:}\Varid{xs})\:\Leftrightarrow\:\Varid{ys}\;\mathbin{\sqsubseteq}\;\Varid{xs}\;\lor\;\langle\, \exists\:\Varid{zs}\: : \:\Varid{ys}\mathrel{=}(\Varid{x}\mathbin{:}\Varid{zs})\: : \:\Varid{zs}\;\mathbin{\sqsubseteq}\;\Varid{xs}\,\rangle}
	\label{eq:pw-sublist}
\end{eqnarray}

Both sublist and prefix are inductive partial orders. While prefix keeps
the head of the (current) list until it decides to stop --- cf. \ensuremath{\Varid{ys}\mathrel{=}[\mskip1.5mu \mskip1.5mu]} in (\ref{eq:pw-prefix})
-- sublist decides either to drop the head or not at each stage of the induction.
This difference will have an impact on the implementation of \ensuremath{\Varid{filter}} when
compared to \ensuremath{\Varid{takeWhile}} --- see \cite{PP23}--, but no impact whatsoever concerning
properties such as e.g.\ (\ref{eq:240528a}), which holds when \ensuremath{\Varid{takeWhile}} is replaced by \ensuremath{\Varid{filter}}.

The accuracy of both formal and informal descriptions of \ensuremath{\Varid{filter}} sharpens again
one's ability to assess its documentation. Clearly, (\ref{eq:240529a}) is
not accurate enough in its allowing for sublist permutations: phrase \emph{``those elements''}
does not ensure they are listed in the original order.\footnote{Elixir's description,
\emph{"\ensuremath{\Varid{filter}\;(\Varid{enumerable},\Varid{fun})} filters the \ensuremath{\Varid{enumerable}}, i.e.\ returns
only those elements for which \ensuremath{\Varid{fun}} returns a truthy value"} \cite{elixir/1.12}
suffers from the same problem. The reader is invited to look for \ensuremath{\Varid{filter}}
descriptions in other languages and check.}

Picking a final example from \cite{PP23}, let us consider:
\begin{quote}\em
\ensuremath{\Varid{dropWhile}\;\Varid{p}\;\Varid{xs}} returns the suffix remaining after \ensuremath{\Varid{takeWhile}\;\Varid{p}\;\Varid{xs}} \cite{base-4.19.1.0}.
\end{quote}
We see that another \emph{concept} is required this time, that of a \emph{suffix}
of a list. Resorting to \ensuremath{\Varid{takeWhile}} is clever and accurate, but property
\ensuremath{\Varid{takeWhile}\;\Varid{p}\;\Varid{xs}\plus \Varid{dropWhile}\;\Varid{p}\;\Varid{xs}\mathrel{=}\Varid{xs}} has to be assumed. To get a definition
of \ensuremath{\Varid{dropWhile}} independent of \ensuremath{\Varid{takeWhile}} another ordering (suffix) has to
be defined on lists, denoted by \ensuremath{(\mathbin{\curlyeqprec})}:
\begin{eqnarray}
\ensuremath{\begin{lcbr}\Varid{s}\;\mathbin{\curlyeqprec}\;[\mskip1.5mu \mskip1.5mu]\:\Leftrightarrow\:\Varid{s}\mathrel{=}[\mskip1.5mu \mskip1.5mu]\\\Varid{s}\;\mathbin{\curlyeqprec}\;(\Varid{h}\mathbin{:}\Varid{t})\:\Leftrightarrow\:\Varid{s}\mathrel{=}(\Varid{h}\mathbin{:}\Varid{t})\;\lor\;\Varid{s}\;\mathbin{\curlyeqprec}\;\Varid{t}\end{lcbr}} \label{eq:pw-suffix}	
\end{eqnarray}
Assuming (\ref{eq:pw-suffix}), one may describe \ensuremath{\Varid{dropWhile}} as follows,
\begin{quote}
\textit{\ensuremath{\Varid{dropWhile}\;\Varid{p}} yields the longest suffix of the input list whose head fails to satisfy predicate \ensuremath{\Varid{p}}}
\end{quote}
leading to
\begin{eqnarray}
\ensuremath{\Varid{headFails}\;\Varid{p}\;\Varid{z}\mathrel{\wedge}\Varid{z}\;\mathbin{\curlyeqprec}\;\Varid{l}\;\,\equiv\,\;\Varid{z}\;\mathbin{\curlyeqprec}\;\Varid{dropWhile}\;\Varid{p}\;\Varid{l}} \label{eq:gc-dropWhile}
\end{eqnarray}
where
\begin{quote}
\begin{hscode}\SaveRestoreHook
\column{B}{@{}>{\hspre}l<{\hspost}@{}}%
\column{19}{@{}>{\hspre}l<{\hspost}@{}}%
\column{E}{@{}>{\hspre}l<{\hspost}@{}}%
\>[B]{}\Varid{headFails}{}\<[19]%
\>[19]{}\mathbin{::}(\Varid{t}{\,\rightarrow\,}\Conid{Bool}){\,\rightarrow\,}[\mskip1.5mu \Varid{t}\mskip1.5mu]{\,\rightarrow\,}\Conid{Bool}{}\<[E]%
\\
\>[B]{}\Varid{headFails}\;\Varid{p}\;[\mskip1.5mu \mskip1.5mu]{}\<[19]%
\>[19]{}\mathrel{=}\Conid{True}{}\<[E]%
\\
\>[B]{}\Varid{headFails}\;\Varid{p}\;(\Varid{h}\mathbin{:}\Varid{t})\mathrel{=}\neg \;\Varid{p}\;\Varid{h}{}\<[E]%
\ColumnHook
\end{hscode}\resethooks
\end{quote}
The calculation of the implementations of \ensuremath{\Varid{dropWhile}} and \ensuremath{\Varid{filter}} from respectively
(\ref{eq:gc-dropWhile}) and (\ref{eq:240530a}) follow the same routine as above and
can be found in \cite{PP23}.

\section{Why Galois connections} \label{sec:240531e}
In its most generality, the \emph{easy-hard-split} design pattern \cite{MO12a} illustrated
above in the specification of Haskell functions can be summarized by the equivalence
\begin{eqnarray}
	\ensuremath{\underbrace{\Varid{f}\;\Varid{y}\leq\Varid{x}}_{\Varid{easy}}} & \equiv & \ensuremath{\underbrace{\Varid{y}\;\mathbin{\preccurlyeq}\;\Varid{g}\;\Varid{x}}_{\Varid{hard}}}
	\label{eq:040120e}
\end{eqnarray}
where \ensuremath{\Varid{f}\mathbin{:}\Conid{A}{\,\rightarrow\,}\Conid{B}} and \ensuremath{\Varid{g}\mathbin{:}\Conid{B}{\,\rightarrow\,}\Conid{A}} are functions and types \ensuremath{\Conid{A}} and \ensuremath{\Conid{B}} come
equipped with two partial orders, respectively \ensuremath{(\mathbin{\preccurlyeq})} and \ensuremath{(\leq)}.\footnote{
Partial orders are not strictly needed, as preorders suffice \cite{Ba04a}. But the use of
indirect equality that comes together with the \emph{easy-hard-split} requires them.
} Wherever
(\ref{eq:040120e}) holds in mathematics, \ensuremath{\Varid{f}} and \ensuremath{\Varid{g}} are said to be ``Galois
connected'' and the equivalence (\ref{eq:040120e}) itself is called a \emph{Galois
connection}, named after Evariste Galois (1811-1832), the French mathematician
who first contributed to the concept. In the case of (\ref{eq:240528d}), for instance,
\ensuremath{\Varid{f}\mathrel{=}\widehat{\Varid{zip}}} and \ensuremath{\Varid{g}\mathrel{=}\Varid{unzip}}, where the hat in \ensuremath{\widehat{\Varid{zip}}} denotes \ensuremath{\Varid{zip}} uncurried.
In the case of \ensuremath{\Varid{take}}, one has \ensuremath{\Varid{f}\mathrel{=}\widehat{\Varid{take}}} and \ensuremath{\Varid{g}\;\Varid{ys}\mathrel{=}(\mathsf{length}\;\Varid{ys},\Varid{ys})}.

The \emph{adjoints} \ensuremath{\Varid{f}} and \ensuremath{\Varid{g}} can be regarded as ``imperfect inverses''
of each other in the sense that the ``round trips''
\ensuremath{\Varid{f}\;(\Varid{g}\;\Varid{x})\leq\Varid{x}} and \ensuremath{\Varid{y}\;\mathbin{\preccurlyeq}\;\Varid{g}\;(\Varid{f}\;\Varid{y})} --- the two cancellations of (\ref{eq:040120e})
--- are not equalities in general. 

Galois connection based reasoning has found many applications in computing,
see e.g.\ references \cite{CC77,Ka98,NNH99,Ba04a,MO12a} among others. They generalize
to adjunctions, which have deepened our understanding of key features in functional
programming, see e.g.\ \cite{Hinze13,Ol23}. The main advantage lies in its rich theory,
which brings with it much economy of thought and effective reasoning.

Just to give a brief example of this richness, let us consider the equalities
\begin{eqnarray}
\start	\ensuremath{\Varid{g}\;(\Varid{f}\;(\Varid{g}\;\Varid{x}))\mathrel{=}\Varid{g}\;\Varid{x}}
	\label{eq:240603a}
\more	\ensuremath{\Varid{f}\;(\Varid{g}\;(\Varid{f}\;\Varid{y}))\mathrel{=}\Varid{f}\;\Varid{y}}
	\label{eq:240531c}
\end{eqnarray}
that can be shown to arise from (\ref{eq:040120e}) for partial orders.
It is also part of the theory that, for \ensuremath{\Varid{f}} injective, \ensuremath{\Varid{g}\;(\Varid{f}\;\Varid{y})\mathrel{=}\Varid{y}} holds \cite{Ol05}.

Let us see what (\ref{eq:240531c}) can offer concerning (\ref{eq:240528d}).
Since we have the definition of \ensuremath{\Varid{unzip}} --- the \emph{easy} adjoint of (\ref{eq:gc-zip})
 --- we can show that it is injective \cite{Ol05}. 
Then (\ref{eq:240531d}) is granted by \ensuremath{\widehat{\Varid{zip}}\;(\Varid{unzip}\;\Varid{xs})\mathrel{=}\Varid{xs}}, even
without knowing the definition of \ensuremath{\Varid{zip}}.

In the same vein, consider the cases of \ensuremath{\Varid{f}} being
\ensuremath{\Varid{takeWhile}\;\Varid{p}} (\ref{eq:240527c}), \ensuremath{\Varid{filter}\;\Varid{p}} (\ref{eq:240530a}) or \ensuremath{\Varid{dropWhile}\;\Varid{p}} (\ref{eq:gc-dropWhile}).
In all these cases \ensuremath{\Varid{f}\;\Varid{ys}\mathrel{=}\Varid{ys}}, and so (\ref{eq:240603a}) becomes \ensuremath{\Varid{g}\;(\Varid{g}\;\Varid{x})\mathrel{=}\Varid{g}\;\Varid{x}}.
A function with this behaviour is said to be \emph{idempotent}. 
We conclude that \ensuremath{\Varid{takeWhile}\;\Varid{p}}, \ensuremath{\Varid{filter}\;\Varid{p}} and \ensuremath{\Varid{dropWhile}\;\Varid{p}} are all idempotent
functions, that is, applying them twice (or more times) is the same as doing it once.
All this, let us once again stress it, is granted by the underlying theory \emph{before}
we know their actual implementations.

\section{Non-Galois connections} \label{sec:240603b}

Just by itself, the lexical meaning of the \emph{``un''} prefix of ``\ensuremath{\Varid{unzip}}''
suggests this function as some kind of inverse of \ensuremath{\Varid{zip}}, which we have confirmed
above, both formally and informally. This may lead one to think that similar
choices of terminology will lead to the same conclusions, but this is not
true. Here are two (counter) examples widely used in string processing in Haskell
--- \ensuremath{\Varid{words}} / \ensuremath{\Varid{unwords}} and \ensuremath{\Varid{lines}} / \ensuremath{\Varid{unlines}}. Informally, both pairs of
functions seem to share the same ``design pattern'': \emph{given a list of
objects of a given type, choose a particular such object to act as separator
in splitting such lists into lists of lists}.

From the outset, everything seems to point to two more Galois connections.
However, closer inspection will reveal that neither is so, despite their
initial appearance.

A simple way to verify this is to go back to (\ref{eq:240603a},\ref{eq:240531c}) and
pick the \ensuremath{\Varid{words}}/\ensuremath{\Varid{unwords}} pair, for instance, where the space character acts as separator.
It is easy to find counter-examples to \ensuremath{\Varid{unwords}\comp \Varid{words}\comp \Varid{unwords}\mathrel{=}\Varid{unwords}}, which by
(\ref{eq:240531c}) should hold. For instance,
	\ensuremath{\Varid{unwords}\;[\mskip1.5mu \text{\ttfamily \char34 Hello~~~\char34}\mskip1.5mu]}
yields
	\ensuremath{\text{\ttfamily \char34 Hello~~~\char34}}, which passed on to \ensuremath{\Varid{words}} gives
        \ensuremath{[\mskip1.5mu \text{\ttfamily \char34 Hello\char34}\mskip1.5mu]}
which the second \ensuremath{\Varid{unwords}} finally converts to \ensuremath{\text{\ttfamily \char34 Hello\char34}}.
If (\ref{eq:240531c}) does not hold for the \ensuremath{\Varid{words}}/\ensuremath{\Varid{unwords}} pair
--- as currently implemented in Haskell ---
they cannot form a Galois connection, i.e.\ they cannot
be an instance of the \emph{easy-hard-split} design pattern.
Similar counter-examples can be easily found for the \ensuremath{\Varid{lines}} / \ensuremath{\Varid{unlines}} pair.
%

\section{Generalization}

In any field of engineering, reaching the best solution as the \emph{right approximation} of certain constraints in some given domain is the hallmark of a good engineer. Such optimization is also at the core operational research. Training a deep neural network can also be considered an approximation process, minimizing classification errors through round-trip layer traversals.

Calculating the right program by narrowing the gap between a loose specification and its best implementation (under some constraints, again) is indeed the essence of textbooks such as e.g.\ \cite{BM97} and permeates the theory of stepwise program refinement \cite{Me09}, for instance. So it seems that viewing programming as the art of reaching \emph{best approximations} \cite{MO12a} is not only pedagogical but also realistic.

It is worth noting that the approach presented in this paper goes beyond list processing, which was the focus of the examples presented above. The prefix and sublist orderings are instances of generic, inductive relations over trees (cf.\ subtrees etc) which can be easily defined in the pointfree style, see e.g.\ \cite{Ol05}. 

\section{Summary} \label{sec:summary}

Software is unquestionably one of the key technologies in modern IT society.
More and more, programming consists of clever \emph{reuse} of code available from
large software libraries and repositories. Knowing exactly what an imported
functionality \emph{actually does} is crucial to software reuse, calling for
clarity and unambiguous specification.  Rigorous specification can be achieved
by use of formal methods, but this requires a background on logic and algebra.
Rarely using formal specs, software documentation uses prose to describe
the meaning of code. Achieving precision in such a documentation requires
a \emph{go-formal-informally} \cite{Mo24} approach to programming based on
solid \emph{concepts} \cite{Ja21}.

This paper addresses a generic design pattern, termed the \emph{easy-hard-split}
\cite{MO12a}, in which prose and maths go together in functional specification.
The role of this pattern is shown through examples taken from the Haskell
Data-List library and similar libraries written in other languages.
The linguistic pattern says how the code is optimal in some sense when performing some particular task.
The underlying maths is the concept of a Galois connection, a rich construction
whereby properties of the specified code are inferred with no need for an \emph{explicit proof}
based on its analysis.

More examples could be given of how much (laborious) proof work is avoided by following this
kind of approach. Almost two centuries ago, this was exactly what worried Evariste Galois himself,
who was quite critical about ineffective reasoning in mathematics:
\begin{quote}\em
``Certaines personnes ont \em [l'affectation] \em d'\'eviter en apparence toute
esp\`ece de {calcul}, en traduisant par des phrases fort longues ce
qui s'exprime tr\`es br\`evement par {l'alg\`ebre}, et ajoutant ainsi \`a
la longueur des op\'erations, les longueurs d'un langage qui n'est pas fait
pour les exprimer.

\par
Ces personnes-l\`a sont en {arri\`ere de cent ans}.''
\\
\hfill{ (1831) }
\end{quote}

\begin{acks}
The authors thank the reviewers of the Haskell Symposium '24 for their suggestions to improve the manuscript.
The second author's work is financed by National Funds through
the FCT - Funda\c c\~ao para a Ci\^encia e a Tecnologia, I.P. (Portuguese
Foundation for Science and Technology) within the IBEX project, with reference
PTDC/CCI-COM/4280/2021.
\end{acks}

\bibliographystyle{ACM-Reference-Format}
\bibliography{galois}

\end{document}